\documentclass[epj]{svjour}

\usepackage{lineno,hyperref}

\newcommand{ \ep }{ $e+p$ }
\newcommand{ \eA }{ $e+$A }
\newcommand{ \dA }{ $d+$Au }

\usepackage{graphicx}
\usepackage{graphics}
\usepackage{epsfig}
\usepackage{epstopdf} 
\usepackage{subfigure}

\usepackage{grffile} 

\begin{document}

\title{Determination of electron-nucleus collision geometry with forward neutrons}

\author{L. Zheng \inst{1,2}
\and E.C. Aschenauer \inst{2}
\and J.H. Lee \inst{2}
}

\institute{
Key Laboratory of Quark and Lepton Physics (MOE) and Institute
of Particle Physics, Central China Normal University, Wuhan 430079, China
\and
Physics Department, Brookhaven National Laboratory, Upton, NY 11973, U.S.A.
}

\date{\\}

\abstract{
There are a large number of physics programs one can explore in electron-nucleus
collisions at a future electron-ion collider. Collision geometry is very
important in these studies, while the measurement for an event-by-event
geometric control is rarely discussed in the prior deep inelastic scattering
experiments off a nucleus. This paper seeks to provide some detailed studies on
the potential of tagging collision geometries through forward neutron
multiplicity measurements with a zero degree calorimeter. This type of geometry
handle, if achieved, can be extremely beneficial in constraining nuclear effects
for the electron-nucleus program at an electron-ion collider.
} 
\maketitle
%

\section{Introduction}

Electron-nucleus ($e+$A) collisions at an electron-ion collider (EIC) provide an
excellent tool for studies to understand the nuclear structure in quantum
chromodynamics (QCD) with its wide kinematic reach and the
ability to accelerate a large variety of nuclear
beams~\cite{Accardi:2012qut}. A wide range of nuclear effects can be
investigated with the EIC facility. For instance, the parton distributions,
especially for gluons at small momentum fraction $x$, are assumed to be largely modified by
the nuclear environment and are still unconstrained. Due to the overlap
of the gluon cloud from different nucleons, the gluon saturation effects may arise
and are thought to be amplified with a nuclear target. At an EIC, with the precise control
of $Q^{2}$ and $x$ possible by measuring the scattered electrons, this nuclear enhanced
saturation effect can be systematically pinned down with measurements such as the
longitudinal strucutre function $F_{L}$~\cite{Albacete:2009fh} and dihadron
correlations~\cite{Dominguez:2011wm}. Other than gluon saturation in the initial
state, the nuclear medium will introduce a modification to the final state color
neutralization and hadronization. The multiplicity ratio measurement $R^{h}_{A}$
can be used to examine the time development of hadronization~\cite{Accardi:2007in}.
 
All these effects are expected to have a strong dependence on the underlying
collision geometries with respect to the nuclear environment. Gluon saturation
is closely related to the impact parameter through the thickness of nuclear
medium. Changes to the hadronization also correlate with the path length of
fast-moving color charges in the cold nuclear medium. However, there has been
little discussion about the characterization of collision geometry in individual
nuclear deep inelastic scattering (DIS) collisions up to now.

In order to characterize the geometry of collisions in proton-nucleus or nucleus-nucleus collisions, quantities
like the number of binary collisions $N_{coll}$, the number of nucleons participating in
binary interactions $N_{part}$ and impact parameter have been extensively
studied with produced particle multiplicities near central rapidity~\cite{Broniowski:2001ei}. This method has been widely used in
the determination of geometries for numerous measurements at the Relativistic Heavy Ion Collider (RHIC) and
the Large Hadron Collider (LHC)~\cite{Aamodt:2010cz,Abelev:2008ab,Back:2002uc}. Unfortunately, in
nuclear DIS studies at the moment, geometric dependence
can only be studied with the variation of nuclear target atomic number $A$,
after averaging over the geometric effect for that given nuclear type.  

In this work, we detail a description of the determination of collision
geometry for \eA using the neutrons emitted at forward rapidities by target
remnant evaporation process. A similar technique has been used for
correcting auto-bias correlations in centrality determination of \dA and $p+$Pb
collisions at RHIC and the LHC~\cite{Adare:2013nff,Toia:2014wia}. We expound the
design of an \eA collision geometry measurement based on simulations from the DPMJET
Monte Carlo (MC) generator~\cite{Roesler:2000he}. Possible applications of this
measurement have been explored. This type of geometry control, if applied to the
observables in \eA program, not only provides an additional handle on nuclear
effects but also simplifies the procedure of estimating systematic errors. Compared to the method of scanning multiple nuclear types, one
only has to deal with the same systematic uncertainties in one nuclear type
instead of worrying about several systematic uncertainties from different
nuclear beams.

The remainder of the paper is organized as follows: in the Sec.~\ref{sec:geoDef}, we introduce the relevant
quantities utilized to describe the collision geometry and illustrate our
strategy for categorizing different geometries. The results of this categorization
are provided in Sec.~\ref{sec:geoCategory}. Possible applications of this
measurement are developed in Sec.~\ref{sec:application}. In the end, we
summarize our methods in Sec.~\ref{sec:summary}.

\section{Characterization method of collision geometry} \label{sec:geoDef}

It has been argued that soft neutron production in lepton-nucleus collisions is
a sensitive probe of final-state interactions in the nuclear
environment~\cite{Strikman:1998cc,White:2010tu}. To our current understanding, such neutrons
are produced in the thermal emission stage of the residual nucleus left after
interactions between the fast probe and target. In the first approximation, the
soft neutron production is proportional to the number of nucleons removed in the
DIS interaction and the subsequent secondary interactions between the
particles generated in DIS process and the rest of the nucleons. In the
following discussions, it is preferable to use the target rest frame with the virtual
photon defining the longitudinal direction.

\begin{figure}
\begin{center}
\includegraphics[width=0.65\columnwidth]{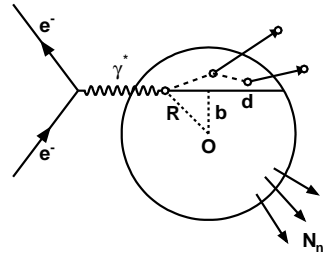}
\caption{Relevant quantities to describe the collision geometry. $b$ represents the impact parameter. $R$ shows the spatial displacement of the interaction point to the center of the nucleus. $d$ is the traveling length, which defines the projected virtual photon traveling length from the interaction point to the edge of the nuclear medium.}
\label{fig:geometry}
\end{center}
\end{figure}

Conventionally, the procedure of an electron scattering from nucleus can be
described by the following steps. With the electromagnetic exchange, the
incoming electron emits a photon, which couples with the nucleus target.
Depending on kinematics, the photon projectile then goes through one or multiple
collisions with partons from the nucleons sitting in the photon's path, which
can be interpreted in various frameworks~\cite{Piller:1995kh}. Here, we define
the length of the projected straight trajectory of the incoming virtual photon
through the nucleus, starting from the involved nucleon during the DIS process
by the traveling length $d$ in the nuclear rest frame, see
Fig.~\ref{fig:geometry}. If there are multiple nucleons involved in the DIS
scattering, $d$ is defined with an interaction point from the average position
of all involved nucleons. Therefore, in each event, one has the position of the
involved nucleon (or average position of multiple nucleons) as the
photon-nucleon primary collision vertex, as presented in
Fig.~\ref{fig:geometry}, at a displacement of $R$ from the center of the nucleus
with an impact parameter $b$.

For large nuclei, a Woods-Saxon distribution is often used to describe the 
initial nuclear density~\cite{Miller:2007ri}:
\begin{equation}
\rho(R)=\frac{\rho_{0}}{1+e^{\frac{R-R_{0}}{a}}},
\label{eqn:woodsaxon}
\end{equation}
in which $R_{0}$ corresponds to the typical nuclear radius $\rho_{0}$ is the nucleon density in the center of the nucleus and $a$ gives the skin depth.

The hadronic fragments generated from photon-nucleon collisions may cause
sequential secondary interactions that knock out additional nucleons
not involved in the DIS interactions. This process is usually named as the
``intranuclear cascade" process~\cite{Bertini:1963zzc}.

After all the formed final-state particles leave the nucleus remnant, due to
momentum conservation, a recoil momentum will lead the residual nucleus to an
equilibrium state characterized by its mass, charge and excitation energy. At the end of the reaction chain, the
excited nucleus will break up into stable final products, with the
emission probability described by the nuclear evaporation model~\cite{Weisskopf:1937zz}. 

In the statistical evaporation model, the number of neutron emissions strongly
depends on the excitation energy, which comes from the number of nucleons removed from
the nuclear remnant, dictated by the successive primary and secondary
interactions in the cascade process~\cite{Ferrari:1995cq}. One may find the
connection between the collision geometry and the evaporated neutron number
distribution through the traveling length $d$ defined above. The larger $d$
is, the more nucleons are expected to be involved in the sequential collisions
and removed from the nuclear remnant, and the more neutrons can be emitted during
the evaporation.

Given such a connection, one may propose that if the emitted
neutron numbers can be measured, we would have a handle to effectively
constrain the underlying collision geometries, which is missing for a long time
in the nuclear DIS studies where the averaged geometry over the whole nucleus
has typically been used. The number of neutrons emitted in the nuclear
break up process will be labeled as $N_{n}$, illustrated in
Fig.~\ref{fig:geometry}.

\begin{figure}[hbt]
\begin{center}
\includegraphics[width=0.55\columnwidth]{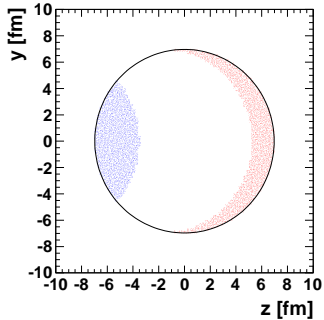}
\caption{A Profile view of the nucleus in the $y-z$ plane in the nuclear rest frame is presented. $y$ shows the impact parameter direction with the virtual photon going in the positive $z$ direction. The radius of the nucleus has been set to $R_{0} =6.52$ fm for gold. The regions showing the position of nucleons involved in the DIS interaction selected by $d <0.25R_{0}$ and $d > 1.5R_{0}$ are marked by red and blue, respectively.}
\label{fig:distBimp}
\end{center}
\end{figure}

Once one can select for traveling length, impact parameter (centrality)
can be effectively constrained according to the traveling length. As is shown
in Fig.~\ref{fig:distBimp}, events with very central collision ($b\approx 0$)
can be acquired with the selection of the largest traveling length (blue
region), although we may have little control on the peripheral collision
events; the red region corresponds to small traveling length, but it
is mixed with central and peripheral collisions.

The Monte Carlo generator DPMJET has been utilized in our studies to simulate
\eA collisions. In the Monte Carlo procedure,
initial nuclear geometric configurations are built with nucleons drawn randomly
with a radius $R$ to the center of target ion from the distribution $4\pi
R^{2}\rho(R)$ where $\rho(R)$ follows the Woods-Saxon distribution. For the
simulation of a gold nucleus, we have $R_{0}=1.12A^{1/3}=6.52$ fm and skin
depth $a=0.545$ fm~\cite{Engel:1996yb}.

The final states in DPMJET are simulated by three stages in a chain. Firstly,
the primary DIS interactions are simulated by PHOJET~\cite{Engel:1995yda}. Only
primary DIS interactions can generate particles with a hard momentum transfer.
Then, particles produced at the primary interaction become the source to trigger the
intranuclear cascade process. A formation zone concept has been introduced to
this cascade process. In the target rest frame of the DIS interaction, a
formation time $\tau$ is needed before newly created particles can re-interact
with the spectator nucleons~\cite{Ferrari:1995cq}:
\begin{equation}
\tau = \tau_{0}\frac{E}{m}\frac{m^{2}}{m^{2}+p^{2}_{\perp}},
\end{equation}
with $\tau_{0}$ being a free formation length parameter. $E$, $m$ and
$p_{\perp}$ are the energy, mass and transverse momentum for the created
particles respectively. For each hadron, a formation time is sampled from an
exponential distribution with an average value as given above. The lower the
hadron energy is, the higher is the probability for that hadron to form inside
the nucleus. The kinematics of the secondary interactions occurring in the
cascade are treated by HADRIN~\cite{Hanssgen:1986az}. Since the nuclear remnant
undergoes equilibration before breakup, evaporative particles should follow a
thermal distribution. Details of the evaporation process are handled by
FLUKA~\cite{Ferrari:2005zk} with the input of remnant charge, mass and
excitation energy and no memory from the prior stages.

It should be noticed that this formation zone intranuclear cascade model is only
one way to describe the effect of final-state interactions in the nuclear
environment. Many other models have more sophisticated considerations with the
prehadron stage or parton energy loss incorporated. Phenomenological studies
have been done by adjusting the prehadron formation time and the final physical
hadron formation time to obtain the best description to the experimental
data~\cite{Akopov:2004ap}. Other QCD-inspired models, which are focused on the
struck quark energy loss, calculate the modifications to the fragmentation
function through gluon bremsstrahlung in a nuclear
medium~\cite{Salgado:2003gb,Chang:2014fba}. No hadron re-interaction is included
in this type of model, as the produced hadron is assumed to always form outside
the nucleus. In this work, we are mainly interested in the correlation between
the underlying geometry and the neutron emission during evaporation. Although
they are bridged through this final state interaction in the nucleus, the exact
details are not very important to our study. We are aware of the possibility to
apply other models like GEMINI~\cite{Charity:1988zz}, SMM~\cite{Bondorf:1995ua}
to treat the nuclear break up process. But as DPMJET is currently the one most
widely used, we concentrate in this paper to illustrate
the potential of the measurement on DPMJET only. More well-developed models,
with better descriptions to the final state interaction, can be added to our
discussions in the future.

Multiple scatterings have been implemented via the Gribov-Glauber
realization~\cite{Shmakov:1988sc} in this MC model. The primary interaction is
sampled from a sequence of independent binary photon-nucleon collisions based on
an elementary photon-nucleon cross section. When we have more than one nucleon
coupled to the DIS interaction stage, the primary interaction point will be
defined as the average position of all the involved nucleons. However,
considering the elementary cross section is very small, the number of binary
collisions happened in one event is most likely to be one. Coherent scattering
effects have been incorporated as a shadowing of the elementary cross section,
based on the coherent length of the photon probe hadronic
fluctuation~\cite{Piller:1999wx}.

\section{Collision geometry categorization} \label{sec:geoCategory}

\subsection{Geometry constraint with forward neutrons}
In the following discussions, we will use the conventional variables to describe
the kinematics in a DIS event. The variable $x=Q^{2}/2M\nu$ is the Bjorken variable, with
$Q^{2}$ being the square of the four-momentum of the exchanged virtual photon,
and $\nu$ showing the virtual photon energy in the target rest frame, if we
denote the nucleon mass as $M$. The variable $y$ shows the inelasticity of the event
measuring the fraction of the electron energy taken by the exchanged boson. The
event sample analyzed in this study is generated from DPMJET for $e+$Au at 10
GeV $\times$ 100 GeV/u with $1 \, \mathrm{GeV}^{2} < Q^{2} < 20 \,
\mathrm{GeV}^{2}$, $ 0.01<y<0.95$.

\begin{figure}
\begin{center}
\includegraphics[width=0.75\columnwidth,keepaspectratio]{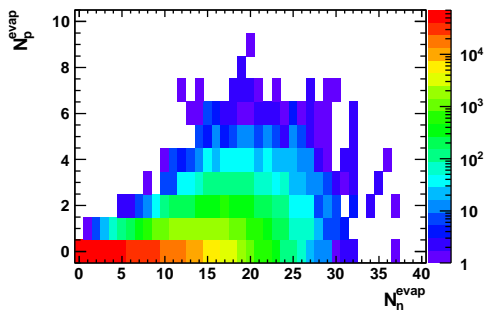}
\caption[proton neutron number correlation during evaporation]{Correlation between the number of evaporated protons ($N^{evap}_{p}$) and neutrons ($N^{evap}_{n}$). Due
to the Coulomb barrier for protons, the proton emission during
evaporation process is greatly suppressed compared to that of neutron.}
\label{fig:evapNeutronVsProton}
\end{center}
\end{figure}

The evaporated products emitted during nuclear break up are most likely to be protons
and neutrons. Due to existence of the Coulomb barrier for charged fragments,
proton emission will be largely suppressed. As seen in
Fig.~\ref{fig:evapNeutronVsProton}, the number of emitted protons during nuclear
evaporation is much lower than that of neutrons from the same event.
Therefore, by measuring neutrons alone, we can characterize the major
properties of the nuclear break up process.

The zero degree calorimeter (ZDC) designed to measure neutral energy deposits
within a small radiation cone about the beam direction can be employed in the
measurement of spectator neutrons emitted with a small angle from the beam
remnant. Meanwhile, charged fragments and the noninteracted
beam remnants will be bent to larger angles out of the ZDC acceptance by
deflecting magnets. Thus, a ZDC reads out the total
neutral energy in the forward rapidities, or effectively the number of
emitted neutrons, from nuclear break up in a very clean way.

\begin{figure}
\begin{center}
\includegraphics[width=0.75\columnwidth,keepaspectratio]{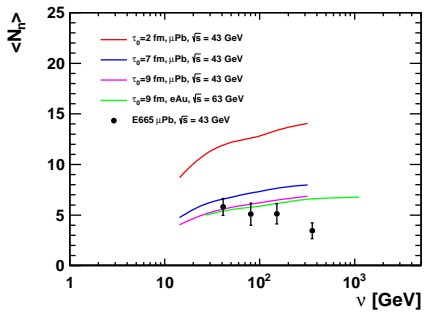}
\caption[comparison with E665]{Results of different formation length parameters as a function of $\nu$ in the simulation compared with the E665 data~\cite{Adams:1995nu}. The green line shows the prospective $e+$Au data kinematics coverage at an EIC.}
\label{fig:tauCompare}
\end{center}
\end{figure}

There's currently only limited knowledge about the magnitude of this neutron emission
for DIS events, the only available measurement being
from E665 experiment~\cite{Adams:1995nu} at FermiLab. From the comparison of
Fig.~\ref{fig:tauCompare}, one can draw an effective formation length $\tau_{0}=9$ fm$/c$
to consistently describe the magnitude of neutron emission. At the planned EIC
energy scale, this measurement can be developed with much better precision and
even wider kinematic range, to achieve a deeper understanding of this
nuclear remnant response.

If well measured, the neutron number can be used as a handle of the collision
geometry. However, since it is impossible to directly measure the number of
neutrons, we will use the energy deposition in a ZDC to extract the neutron
number information. In the following discussions, an energy resolution $\sigma/E
= (85/\sqrt{E} + 9.1)\%$ from Ref.~\cite{Adler:2000bd} has been used as the ZDC
responses for neutron energy. The resolution of the traveling length is
dominated by its intrinsic correlation with the number of emitted neutrons
during the evaporation process. The assumed ZDC energy resolution, which is used
in the current RHIC heavy ion experiments, is sufficient for our study. Based on
the ZDC performance during RHIC running over a wide range of
energies, an efficiency close to $100\%$ to measure the forward neutrons with a negligible
background has been assumed for all the studies in this paper.

\begin{figure*}[hbt!]
\begin{center} 
\subfigure[]{ 
\centering
\includegraphics[width=0.75\columnwidth,keepaspectratio]{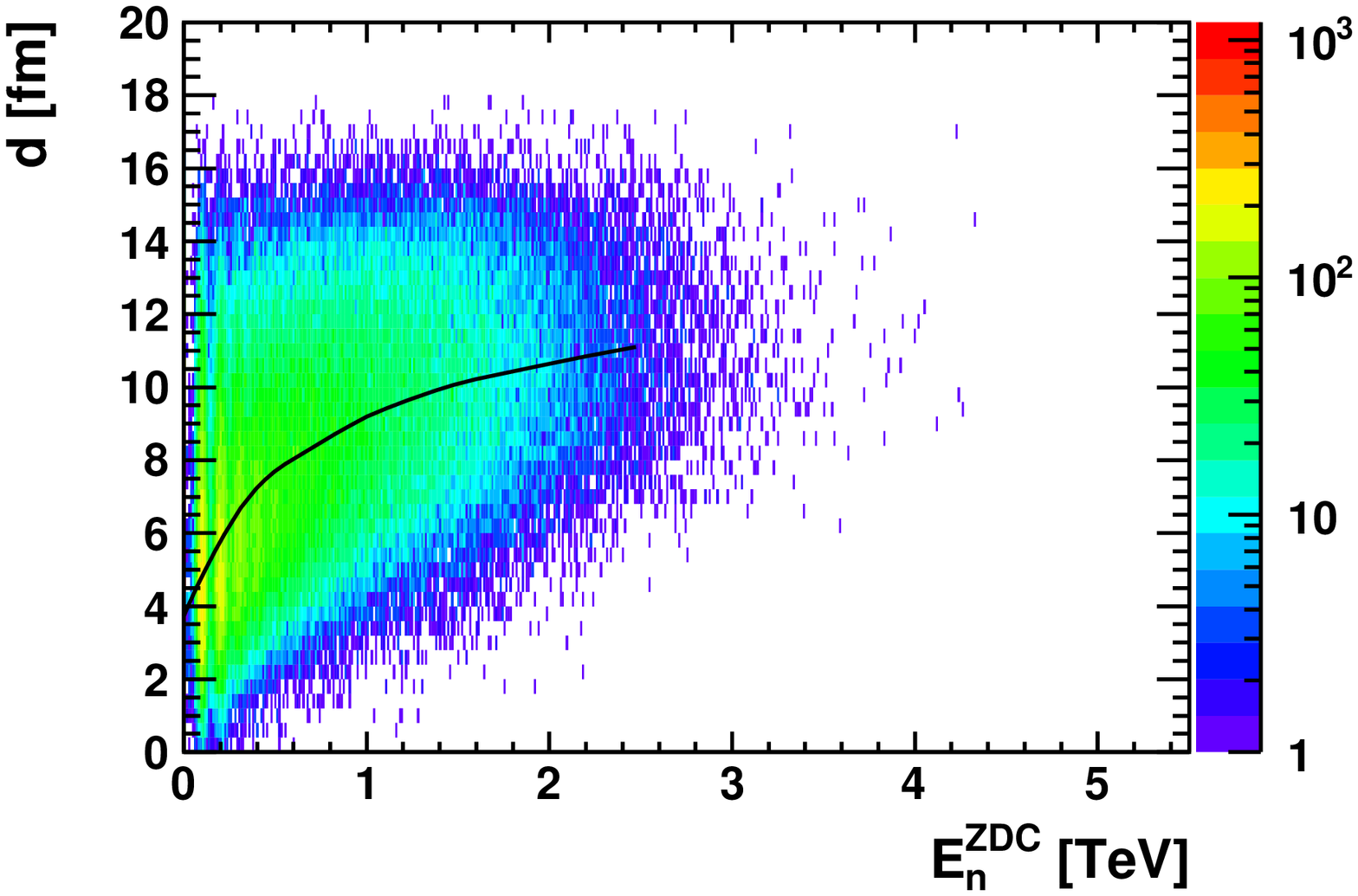} \label{fig:distCorre} 
}
\quad
\subfigure[]{ 
\centering
\includegraphics[width=0.75\columnwidth,keepaspectratio]{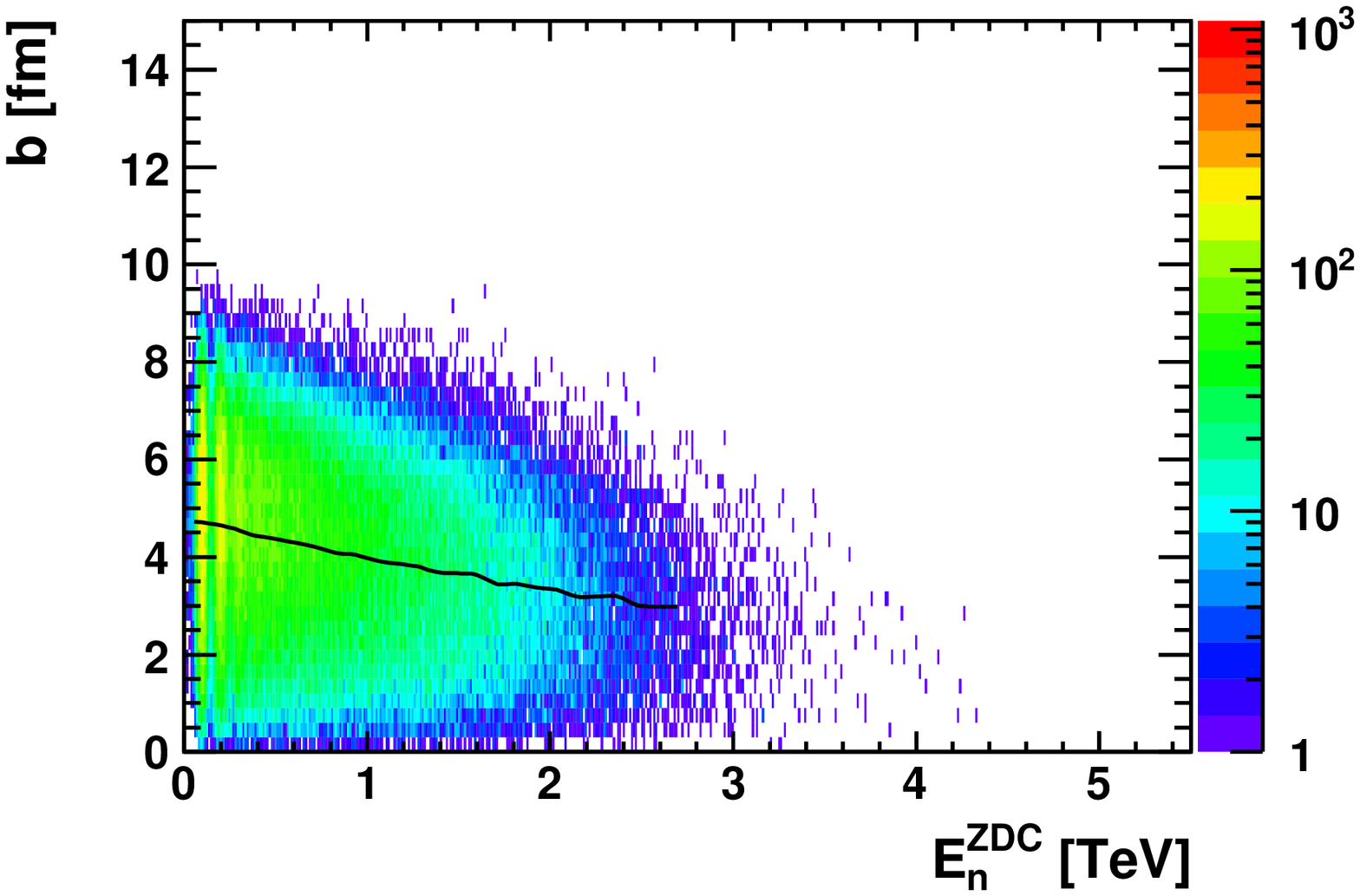} \label{fig:bimpCorre} 
}
\caption[geometry quantity correlation with neutron energy deposit]
{(a) Scatter plot of traveling length $d$ and the energy deposition in the ZDC. The central value of $d$ in every $E_{ZDC}$ bin is indicated by the black line. (b) Scatter plot of the impact parameter $b$ and energy deposition in the ZDC. The central value of $b$ in every $E_{ZDC}$ bin is indicated by the black line.} \label{fig:geoCorrelation}
\end{center} 
\end{figure*}

\begin{figure} 
\begin{center} 
\includegraphics[width=0.75\columnwidth,keepaspectratio]{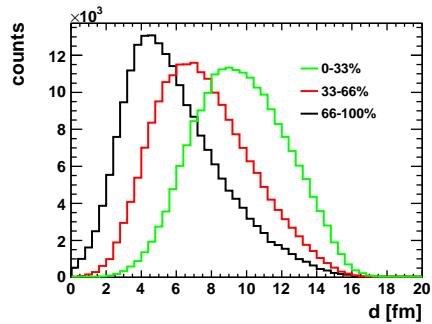}
\caption[geometry quantity constrained by binning method]
{Traveling length distribution in different forward neutron energy bins. The black line corresponds to peripheral collisions (66-100\%), while
the red and green lines correspond to the 33-66\% and 0-33\%, respectively.}
\label{fig:geoConstrain}
\end{center} 
\end{figure}

\begin{figure}
\begin{center}
\includegraphics[width=0.75\columnwidth,keepaspectratio]{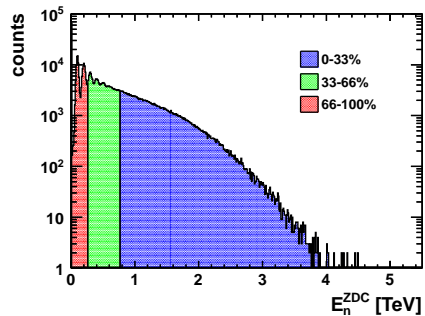}
\caption[neutron binning method]{Selections of different collision geometry with neutron number translated to energy deposition. Different colors represent different centrality selections}
\label{fig:neutronBin}
\end{center}
\end{figure}

\begin{table*}[width=0.85\columnwidth]
\centering
\begin{tabular}{ |l || l | l | l | } \hline 
		& $\Sigma E_{n}$ range [GeV] & $<d>$ / $d_{\mathrm{RMS}}$ [fm] & $<b>$ / $b_{\mathrm{RMS}}$ [fm] 	\\ \hline
0-33\%	& 743-4329		&	9.7 / 2.7	& 3.8 / 1.6	 \\ \hline
33-66\%	& 237-743		&	7.5 / 2.8	& 4.4 / 1.7	\\ \hline
66-100\%& 0-237			&	5.9 / 2.8	& 4.7 / 1.8	\\ \hline
\end{tabular}
\caption[constrained geometry]{Collision geometry constrained by the selection of the neutron energy deposition in the ZDC with the method suggested in Fig.~\ref{fig:neutronBin} . The average value as well as the root mean square (RMS) for the traveling length $d$ and the impact parameter $b$ are presented in this table.}
\label{tab:geoConstr}
\end{table*}

Indicated by Fig.~\ref{fig:geoCorrelation}, energy deposition in the ZDC can be used
as a good measure of traveling length $d$ while the impact parameter $b$ is not as
well controlled. 
Only the most central events ($b\sim0$) can be selected with the largest
evaporated neutron emissions. With this correlation shown in
Fig.~\ref{fig:distCorre}, one can select a binning method to constrain the
underlying geometries. Fig.~\ref{fig:geoConstrain} shows to what extent the
traveling length $d$ can be constrained with the binning method shown in
Fig.~\ref{fig:neutronBin}. The percentage is defined by the fraction of events with
a certain energy deposition in the ZDC.
Constraints put on the quantities with statistical uncertainties under the
current binning strategy can be found in Tab.~\ref{tab:geoConstr}.

\begin{figure}
\begin{center}
\includegraphics[width=0.75\columnwidth]{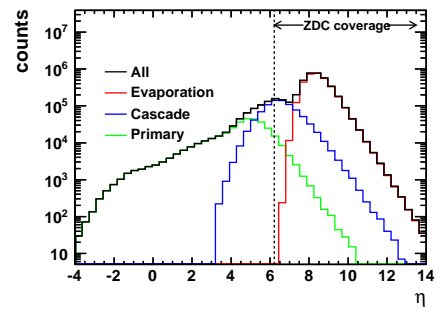}
\caption{$\eta$ distributions of final state neutrons from various processes. Most of the evaporated neutrons can be accepted
if the ZDC covers the polar angle range of $\pm$ 4 mrad marked by the
dashed vertical line. Black lines represent all the final state
neutrons, while the red, blue and green lines illustrate neutron distributions
from evaporated, intranuclear cascade and primary productions.}
\label{fig:neutronZDC}
\end{center}
\end{figure}

The assumed polar angular acceptance of the ZDC is $\pm$ 4 mrad with respect to
the gold nuclear beam direction~\cite{Accardi:2012qut}.
Fig.~\ref{fig:neutronZDC} shows that emitted neutrons from the evaporation
process can be $100\%$ accepted. The final state neutrons from all processes are
marked by the black line in that plot. The green line shows neutrons from
primary interactions. Neutrons generated by intranuclear cascade are shown in
blue and evaporated neutrons are shown in red. It can be concluded from this
plot that neutrons from primary interactions are mainly in the midrapidity
region, while the forward region neutrons are dominated by the evaporated
neutrons. The primary interaction and intranuclear cascade process can also
become a source for final state neutrons accepted by the ZDC. In the simulated
event sample $14.66\%$ of the accepted neutrons in the ZDC come from processes
like primary and secondary interactions. As to the expected detector design, all
the evaporated neutrons can be fully accepted by the experimental device and
contamination from other process is under control.

\subsection{Possible constraint on the most central collisions with a double cut method}

\begin{figure}
\begin{center}
\includegraphics[width=0.75\columnwidth]{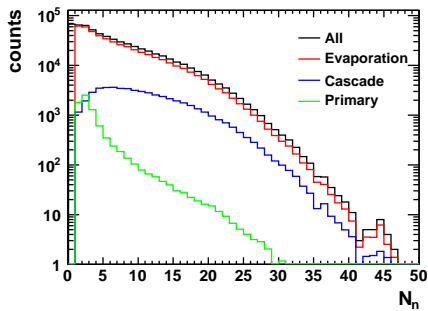}
\caption[accepted neutrons from different stages]{Number of neutron distributions
from different interaction stages. The black line shows the total number of neutrons
within the ZDC acceptance; neutrons from the thermal evaporation process are depicted
by the red line, while the blue line gives the distribution from secondary
interactions and particles generated in primary interactions are marked in green.}
\label{fig:neutronStage}
\end{center}
\end{figure}

The features of particle yield at different stages of the nuclear response
are particle-type-dependent. As for the neutrons, they are mostly generated
in the evaporation process from the ``cooling" of a large excited nucleus.
As shown in Fig.~\ref{fig:neutronStage}, the ZDC accepted
neutrons are mostly evaporated neutrons. By measuring the most forward
neutral energy deposition in the ZDC we can extract the statistical emission
for that event in a very clean way.

With the intranuclear cascade process, additional particles like pions
will be generated in the reaction chain between the fast moving particles
and the rest of the nucleons.
As the $p_{T}$ kick from this type of reaction is very small, charged pions
generated in this process will move to the more forward direction compared
to those pions from primary interactions.

\begin{figure*} 
\begin{center} 
\subfigure[]{ 
\centering
\includegraphics[width=0.75\columnwidth,keepaspectratio]{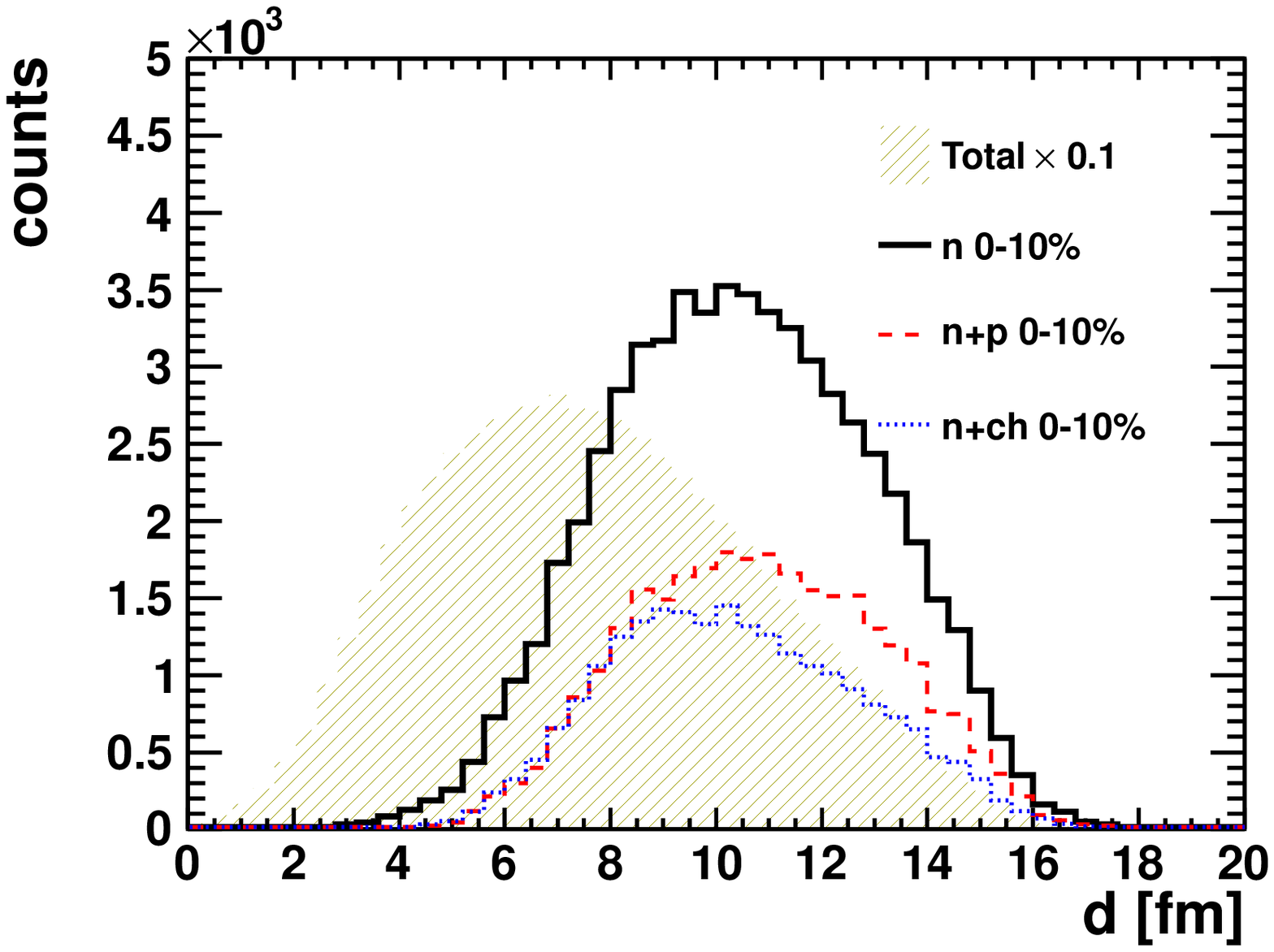} \label{fig:travelConstrain_central} 
}
\quad
\subfigure[]{ 
\centering
\includegraphics[width=0.75\columnwidth,keepaspectratio]{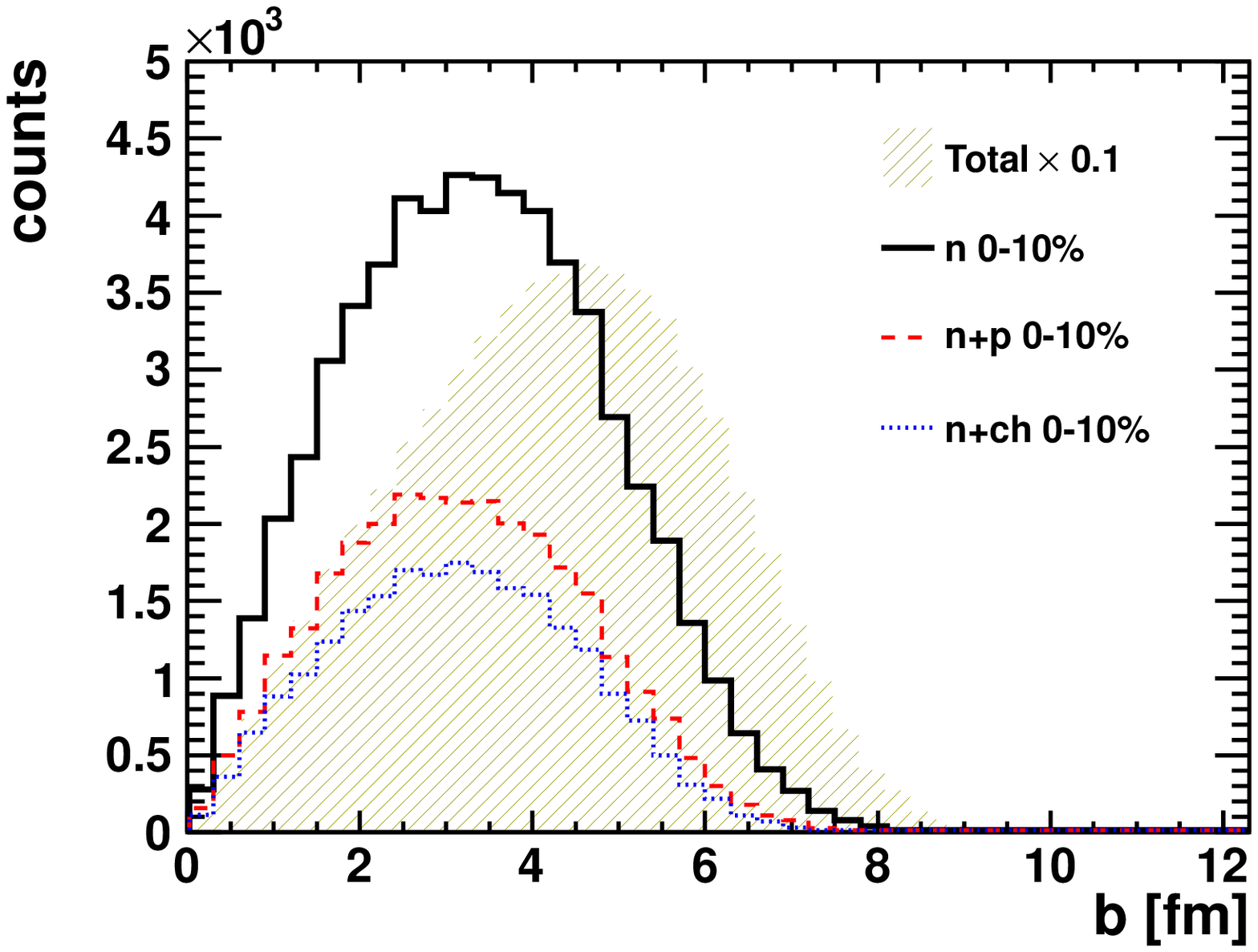} \label{fig:bimpConstrain_central} 
}
\caption
{Binning constraint on traveling length $d$ and impact parameter $b$ for most
central collisions (0-10\%), with forward neutron cut only (solid line) and
neutron cut plus additional forward proton number cut (dashed line) or forward
charged hadron number cut (dotted line). The total distribution has been marked by
the shaded region with a magnitude scaled down by a factor 10 for illustration.
}
\label{fig:geoConstrain_central}
\end{center} 
\end{figure*}

With the knowledge of the underlying traveling length for the collision geometry,
it is possible to define the most central collision by selecting events with the
largest neutron production, in which the incoming photon probes the densest area
of the target. The forward proton track number from break up and the charged
particle (mostly pions) yield in forward rapidities are also sensitive to the
collision geometries. We will see how much we can gain as a constraint on
most central collisions if we put an additional cut on these quantities together
with that on the forward neutrons. It may be beneficial to add the large neutron emission cut
and the large proton or charged particle production cut at forward rapidity at
the same time to select the extremely central bins. Forward proton number is
measured with a perfect resolution in an assumed coverage the same as ZDC.
Forward charged particles are supposed to be counted by devices covering
$4<\eta<6$. The geometry constraint in the most central collisions is
shown in Fig.~\ref{fig:geoConstrain_central}. The shaded region displays the total
distribution for $d$ or $b$ without any cuts. To compare with the cuts made by
0-10\% selection, the magnitude of total distribution has be rescaled by 0.1 for the purpose of
demonstration. Comparing the solid line and shaded region in
Fig.~\ref{fig:geoConstrain_central} shows the forward neutron cut effectively finds
the events with most central collisions. No significant enhancement can be found
with the inclusion of the double cut method from forward proton or charged
particles. Such ZDC accepted neutron energy would be enough to select
the most central events.

\section{Applications of collision geometry constraint at an EIC}\label{sec:application}

As both initial and final state effects in \eA collisions are highly dependent
on nuclear geometry, applying this nuclear geometry handle to our
measurements in \eA program at an EIC will allow us to learn more
about the nuclear medium effect. For instance, the selection of very
central collisions maximizes the probed gluon density in the initial state which
delivers more sensitivity to the expected saturation effect. $F_{L}$ and dihadron
correlation measurements are two important observables sensitive to this
geometry constraint. We would see stronger saturation effects in the most
central collisions compared to peripheral \eA events with the change of
gluon densities. This collision geometry also provides an additional dimension to
the study of final state effects. The time space feature of this fragmentation
process can be directly confronted with the nuclear medium geometry through
which one can explore the hadronization process in a more precise way.

\subsection{Energy loss measurement in the cold nuclear medium}
It has been argued that the multiplicity ratio measurement $R^{h}_{A}$ from
HERMES suggests that the space-time development of the hadronization process in
the nuclear medium can be studied with the semi-inclusive deep-inelastic
scattering process off
nuclei~\cite{Airapetian:2003mi,Airapetian:2007vu,Airapetian:2011jp}. We observe
the final state hadrons with characteristic kinematic variables, such as the
transverse momentum $p_{T}$ and the fractional energy of the
hadron relative to the virtual photon $z$. Depending on the kinematic variables, the
hadron formation may take place inside the nucleus, or outside the nucleus. In
general it is assumed that the struck quark forms its full hadron identity with
an average formation length $l_{h}\propto f(z)\nu$.

The multiplicity ratio $R^{h}_{A}$ is a frequently studied quantity
which effectively describes the nuclear medium energy loss effect. It is defined as follows:
\begin{equation}
R^{h}_{A}=\frac{(\frac{N_{h}(\nu,Q^{2},p^{2}_{T},z)}{N_{l}(\nu,Q^{2})})_{A}}{(\frac{N_{h}(\nu,Q^{2},p^{2}_{T},z)}{N_{l}(\nu,Q^{2})})_{D}},
\end{equation}
where $N_{h}$ is the semi-inclusive production of hadrons in a given kinematics
bin and $N_{l}$ is the yield of leptons in the same $\nu,Q^{2}$ bin. This ratio
is defined for the hadrons production per deep-inelastic scattering event on a
nuclear target with mass $A$ to that of the lightest isoscalar nuclear target
deuterium.

\begin{figure}
\begin{center}
\includegraphics[width=0.75\columnwidth,keepaspectratio]{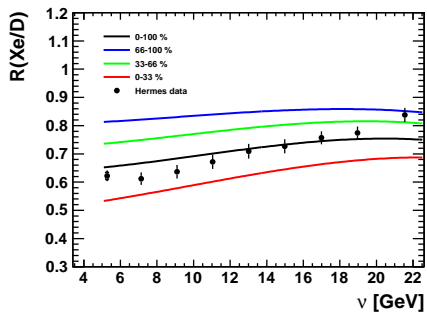}
\caption{Multiplicity ratio as a function of $\nu$ with the traveling length constraint. The black line represents the expected value for minimum bias events, and projected result from three centrality bins are shown in blue, green and red according to our selection discussed in Sect.~\ref{sec:geoCategory}. Hermes data points~\cite{Airapetian:2007vu} are shown in the plot with black dots.}
\label{fig:energyLoss}
\end{center}
\end{figure}

With a handle to constrain the traveling length in the
collision geometry, it is possible to explore the interplay between color
object neutralization and the nuclear environment geometry. If it is possible to
control the traveling length $d$ of the struck quark, one has an additional
dimension in this measurement which changes the measurement from
\(N_{h}(\nu,Q^{2}p^{2}_{T},z)\) to \(N_{h}(\nu,Q^{2}p^{2}_{T},z,d) \).
Instead of varying the different nuclear types, the hadron formation length can be
directly confronted with a traveling length $d$; if $d>l_{h}$ it is generally
formed inside the nucleus and the hadron yield will be suppressed while for
$d<l_{h}$ the medium modification is supposed to be small. With the bin
selection of collision geometry, we have extracted the $d$ distribution from
these bins as input to an energy loss model~\cite{Salgado:2003gb}. Thereby,
strong nuclear medium dependence is expected in the multiplicity ratio
measurement, shown in Fig.~\ref{fig:energyLoss}.

\subsection{Dihadron correlation measurements to probe gluon saturation}

In the small $x$ limit, gluon density in the nucleon becomes so large that a
phenomena named saturation may emerge. It is suggested in this scenario that, when the probe
resolution $Q^{2}$ is less than the saturation scale $Q^{2}_{s}$, the gluon
recombination mechanism becomes dominant, which tames the rapid growth of
gluon density at small $x$. According to the collision
geometry, stronger saturation is expected for the most central collisions. If
this type of physics exists, evaporated neutron number can be used as an
additional handle, together with the kinematics control, for studying the saturation effect.

Dihadron correlations in \eA are a way to investigate the saturation effect. By
selecting the trigger associate particle pairs with certain $p_{T}$, one can
explore the underlying jet properties and the interplay with the saturation
scale $Q_{s}$. The dihadron correlation function can be constructed with the azimuthal angle difference, $\Delta\phi$, between the trigger and associate particle in the same event:
\begin{equation}
C(\Delta\phi)=\frac{N_{pair}(\Delta\phi)}{N_{trig}},
\end{equation}
where $N_{pair}(\Delta\phi)$ is the number of correlated hadron pairs in every
$\Delta\phi$ bin and $N_{trig}$ shows the total number of trigger particles in
the event sample. The saturation scale can be parameterized as a scaling model,
 $Q_{s}^{2}=A^{1/3} c(b)(\frac{x}{x_{0}})^{-\lambda}$, with
$x_{0}=3.04\time 10^{-4}, \lambda = 0.288$. $c(b)$ defines the thickness of the
nuclear medium at impact parameter $b$. Significant saturation effects arise
when $Q_{s}$ is large, which means collisions with smaller $x$ or larger
thickness favor a stronger saturation effect.

Following the formalism in~\cite{Zheng:2014vka}, one can estimate the centrality
dependence in \eA dihadron correlations, which is coordinated by the parameter $c(b)$.
From the simulation based on our centrality definition above, we can select the
events with the most central and most peripheral collisions. Observable suppression
is expected from peripheral to central events.

\begin{figure*}[hbt!]
\begin{center}
\subfigure[]{ 
\centering
\includegraphics[width=0.75\columnwidth,keepaspectratio]{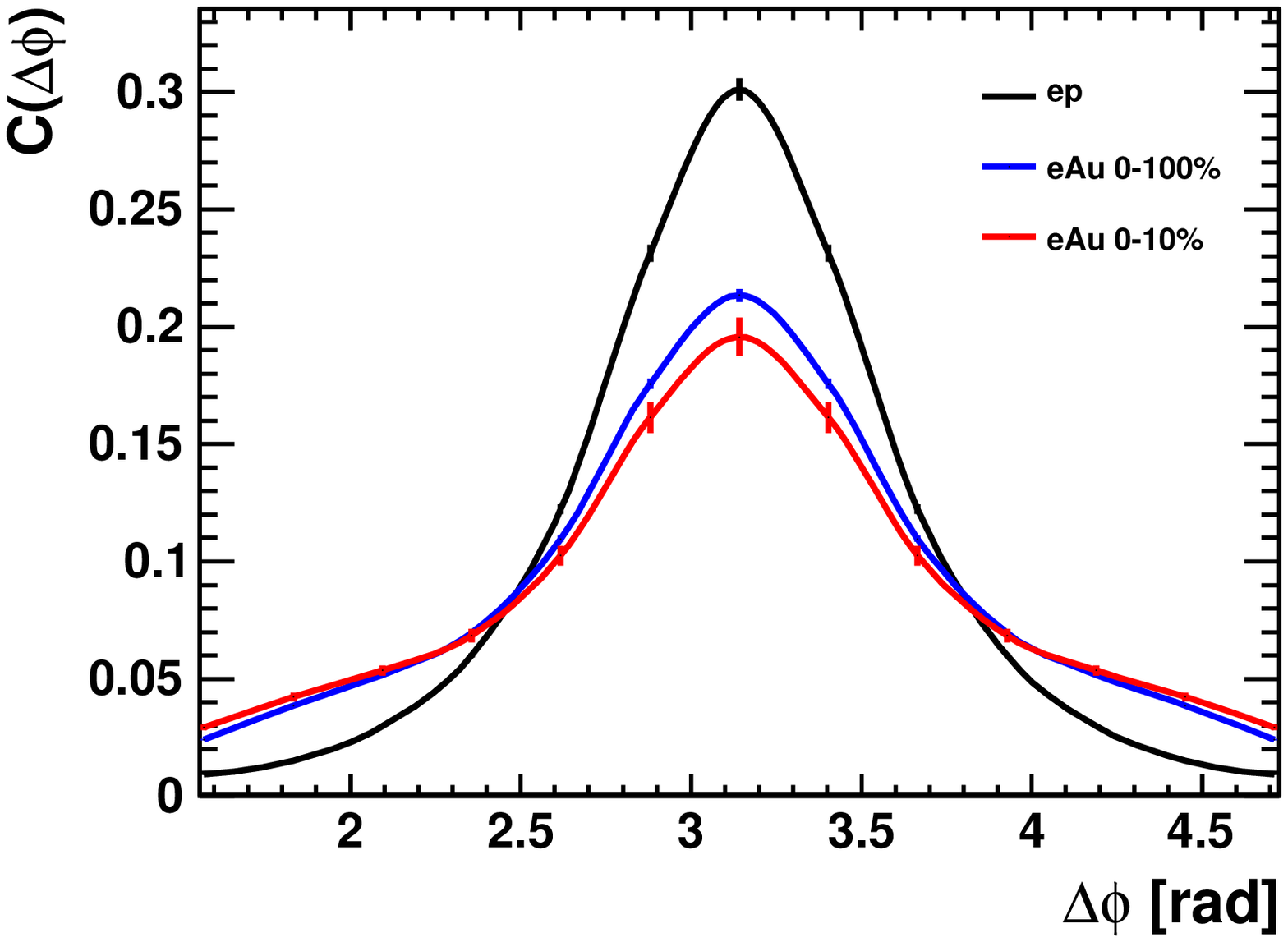}
}
\quad
\subfigure[]{ 
\centering
\includegraphics[width=0.75\columnwidth,keepaspectratio]{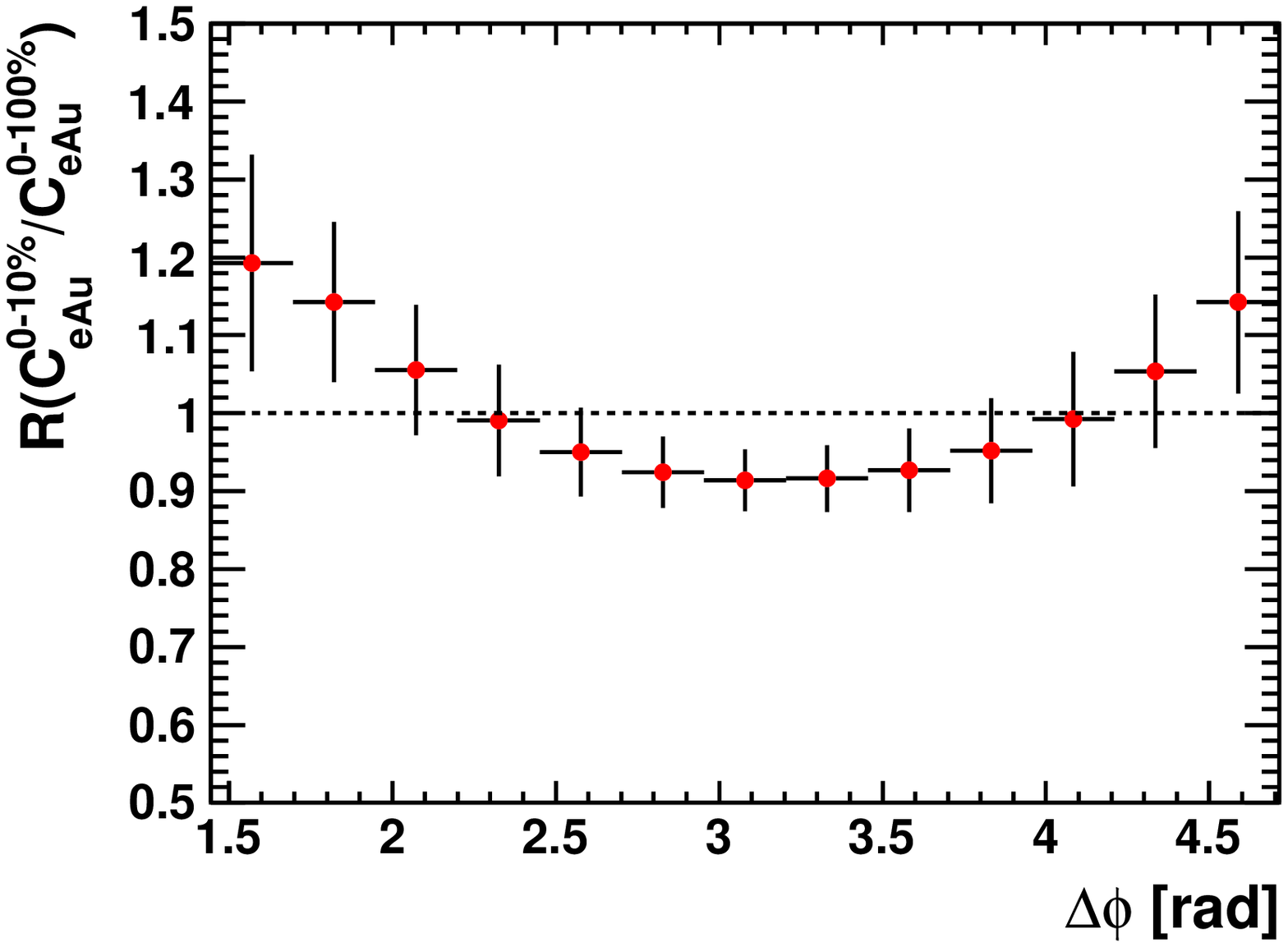} \label{fig:dihadron_ratio} 
}

\caption{Saturation effects in the dihadron correlation measurements with the most central collisions. (a) The correlation functions for \ep, $e+$Au 0-100\% and $e+$Au 0-10\% at 20 GeV$\times$100 GeV are shown in black, blue and red curves with the statistical uncertainty from an integrated luminosity of 10 fb$^{-1}$; (b) the ratio of the correlation function for $e+$Au 0-10\% divided by 0-100\%.}
\label{fig:dihadron_centrality}
\end{center}
\end{figure*}

Fig.~\ref{fig:dihadron_centrality} shows the predicted saturation impact on dihadron correlations with an integrated luminosity of $10fb^{-1}$ at the energy of $\sqrt{s}=63$ GeV.
The effect of selecting the most central collisions has been studied. The thickness function is estimated by
a Woods-Saxon density as in Eq.~\ref{eqn:woodsaxon}. For 0-10\% centrality $c(b)=0.75$, while $c(b)=0.59$
for a minimum bias (0-100\%) estimation. With the current statistical uncertainty, the most central category can be slightly distinguished from the minimum bias collisions. If the ratio of correlation function from central divided by minimum bias is less than 1, as suggested by Fig.~\ref{fig:dihadron_ratio}, it could be an indication of saturation.

\section{Summary} \label{sec:summary}

We have presented detailed studies on the determination of collision geometry
in \eA collisions. Utilizing the DPMJET3 MC generator, we have found a
correlation between the traveling length and the number of neutrons evaporated
from the nuclear remnant. This neutron number distribution can be measured with
a ZDC in the Au-going direction with the systematics under control. All the
devices needed for this measurement have been included in the current EIC
detector design. Therefore, it is very easy to acquire this measurement
with little additional investment. Constraining collision geometry
quantities like traveling length is very meaningful in the studies of cold
nuclear medium effects. A demonstration of using this approach to make
fine traveling length binning of performing nuclear medium energy loss studies
and explore the signature of saturation physics has been provided. With the
determination of collision geometry in these measurements, our understanding of 
nuclear structure can be constrained with higher precision.

\begin{acknowledgement}
We wish to thank Bo-Wen Xiao for helpful discussions and suggestions. This work
was supported in part by the NSFC (11475068), the National Basic Research
Program of China (2013CB837803), and the Basic Research Program of CCNU
(CCNU13F026). E. C. A. and J. H. L. acknowledge support by the U.S. Department
of Energy under Contract No. DE-AC02-98CH10886.
\end{acknowledgement}

\bibliography{reference}{}
\bibliographystyle{bibstyle/epjstyle}

\end{document}